\documentclass[final,5p,times,twocolumn,times]{elsarticle}
\usepackage{amssymb}
\usepackage{graphicx}
\usepackage{xcolor,amsmath,amsfonts,graphicx,amssymb,subfigure}
\usepackage{dblfloatfix}
\usepackage[utf8]{inputenc}
\usepackage{mathtools}

\graphicspath{{./figures/}}

\DeclarePairedDelimiter\ket{\lvert}{\rangle}

\begin{document}

\begin{frontmatter}
\title{Fate of the $\eta'$ in the Quark Gluon Plasma}
\author[Moscow1,Moscow2,Moscow3]{Andrey Yu. Kotov}
\ead{andrey.kotov@phystech.edu}
\author[Florence]{Maria Paola Lombardo}
\ead{lombardo@fi.infn.it}
\author[Samara]{Anton M. Trunin}
\ead{amtrnn@gmail.com}
\address[Moscow1]{Moscow Institute of Physics and Technology, Institutsky lane 9, Dolgoprudny, Moscow region, 141700 Russia}
\address[Moscow2]{Institute for Theoretical and Experimental Physics NRC ``Kurchatov Institute'', Moscow, 117218 Russia}
\address[Moscow3]{Bogoliubov Laboratory of Theoretical Physics, Joint Institute for Nuclear Research, Dubna, 141980 Russia}
\address[Florence]{INFN, Sezione di Firenze, 50019 Sesto Fiorentino (FI), Italy}
\address[Samara]{Samara National Research University, Samara, 443086 Russia}

\begin{abstract}

In this paper we study  the $\eta'$  in $N_f=2+1+1$ lattice QCD simulations at finite temperature. Results are obtained from the analysis of the gluonic defined topological charge density correlator after gradient flow. 
Our results indicate the growth of the $\eta'$ mass above the pseudocritical temperature, associated with the chiral symmetry restoration. In the vicinity of the pseudocritical temperature the results are consistent with a small dip in the $\eta'$ mass. The magnitude of the dip is compatible with the reduction of the $\eta'$ mass obtained by experimental analysis and suggests that $\eta'$ mass comes close to zero temperature non-anomalous contribution.

\end{abstract}

\begin{keyword}
$\eta'$ meson, chiral symmetry, nonzero temperature, lattice QCD,  QCD topology, Quark Gluon Plasma.
\end{keyword}

\end{frontmatter}

\section{Introduction}
\label{bib}

The realization of the chiral and axial symmetries in QCD has important 
phenomenological implications. 
The spectrum of mesons built by light quarks -- up, down, strange --
is nicely accounted for by considering the
spontaneous breaking of chiral symmetry. The well known puzzle associated
with a too heavy $\eta'$ mass found a solution once the non-trivial
topological structure of the vacuum is taken into 
account~\cite{tHooft:1976rip}. The
solution  can be nicely formalized
within the framework of large-$N$ QCD where 
it is shown that the mass matrix should include
a term with the topological susceptibility. Then, the
main features of the physical
spectrum can be reproduced if the topological susceptibility is 
non-zero~\cite{Witten:1979vv,Veneziano:1979ec,DiVecchia:2017xpu}.
These are eminently nonperturbative phenomena, posing specific
challenges: a numerical approach is mandatory, and early lattice studies
have indeed confirmed a non-zero topological susceptibility~\cite{DiVecchia:1981aev},
providing a clear evidence for the validity of 
the Witten-Veneziano analysis~\cite{Cichy:2015jra}.
Moreover, the $\eta'$ mass itself  has been directly measured on the lattice,
with results well in agreement with experimental data~\cite{Michael:2013gka,Ottnad:2017bjt}.

A natural question then arises, concerning the fate of the symmetry patterns
at high temperatures. The study of sequential restoration, or lack thereof,
of the relevant symmetries helps understanding the fundamental mechanisms
underlying these phenomena. 
The restoration of the chiral $SU(2)_{L} \times SU(2)_{R}$
symmetry at high temperatures has been the subject of detailed lattice studies:
there is now consensus that such symmetry is restored in the chiral limit
at a temperature of about $140$ MeV, and that the explicit breaking associated
with the quark masses turns the phase transition into a crossover which occurs
at about $157$ MeV for physical quark masses~\cite{DElia:2018fjp}.  

The fate of the $U(1)_A$ symmetry is a subtler issue, since the $U(1)_A$ anomaly
provides a mechanism for explicit symmetry breaking~\cite{tHooft:1976rip},
which, being at operator level, exists independent
of temperature. However, instantons and their suppression may provide
a mechanism for its effective restoration~\cite{Shuryak:1993ee},
raising the issue of the ordering of the chiral and axial symmetry
restoration.  
Current results are inconclusive: some results indicate a coincidence of the
$U(1)_A$ restoration with that of the chiral restoration, others suggest
that $U_A(1)$ breaking persists up to large temperatures being effectively
restored only in the high temperatures, dilute instanton gas limit. 

The pattern of symmetries at high temperature has of course 
influence on the meson spectrum in the plasma: 
without breaking, the light flavor
pseudo-Goldstone bosons would become (nearly) degenerate  
with their chiral partners.  Concerning the $\eta'$, assuming a restoration
of the $U(1)_A$ symmetry coincident with that of the chiral symmetry,
the natural conclusion would be that $\eta'$ follows the fate of its
chiral partners. The details are subtle and will be reviewed in the
following, and in a short summary ab initio calculations are mandatory
to reach firm conclusions. 

On the experimental side, any variation of the 
mass of the $\eta'$ should produce interesting signatures in 
heavy ions collisions experiments~\cite{Kapusta:1995ww}. 
In Ref.~\cite{Kapusta:1995ww} the authors 
argue that a small mass of the $\eta{'}$
in hot and dense matter should lead to an increase of 
the production cross section with respect to the one in pp.
The most popular experimental results are those from PHENIX
and STAR at 200 GEV gold-gold collisions~\cite{Csorgo:2009pa,Csorgo:2010hj}.
They determine the best value for the in-medium
mass reduction according to their model, suggesting a mass reduction
of about 200 MeV.  In the same work it is  also noted that
different initial abundances will change the results. 
Indeed, one can never directly measure the $\eta'$ mass,
only relative abundances with inherent
ambiguities in the interpretations of the results. 

\begin{figure*}[b]
\begin{center}
\includegraphics[width=9cm,angle=0]{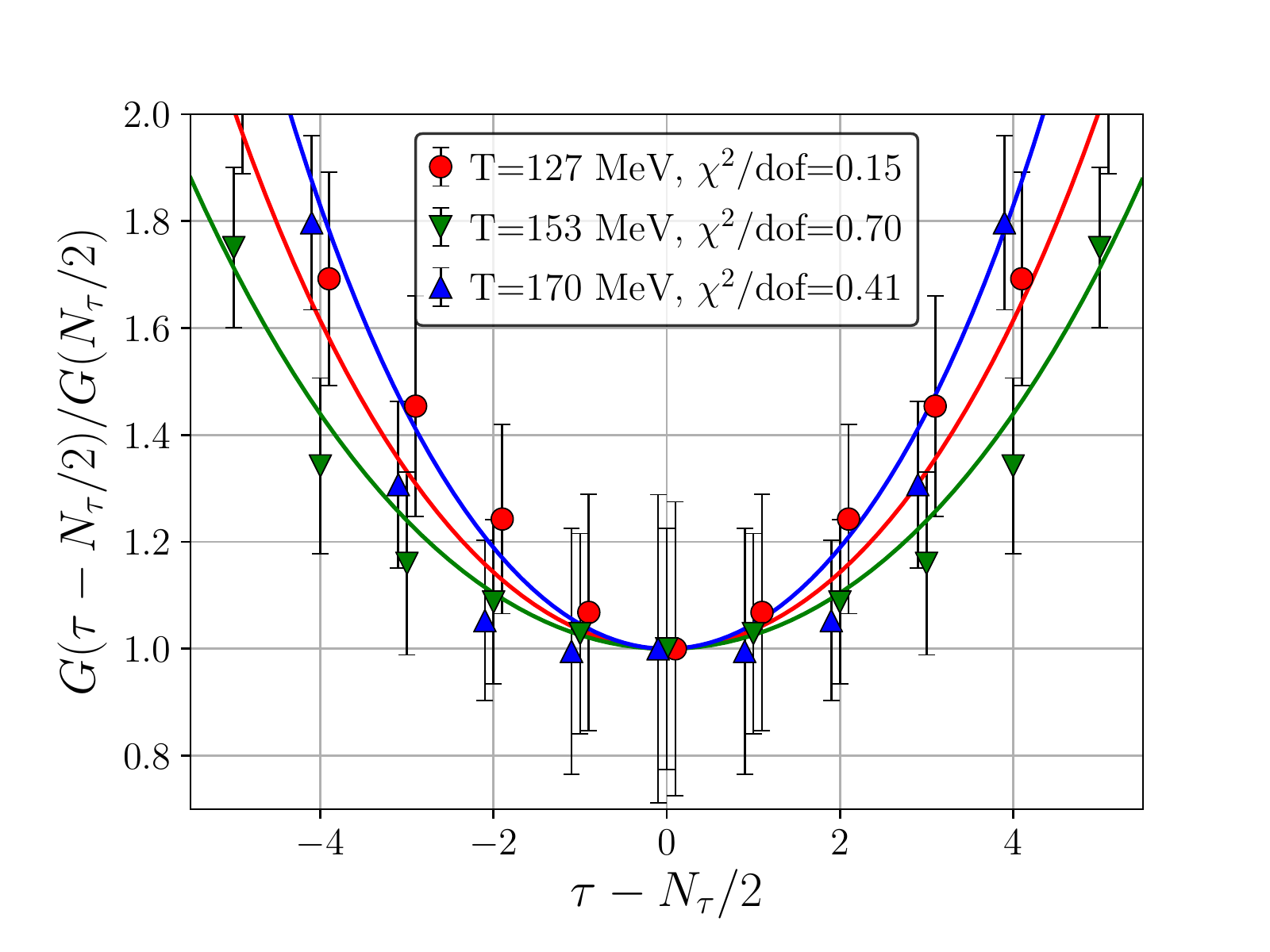}
\includegraphics[width=9cm,angle=0]{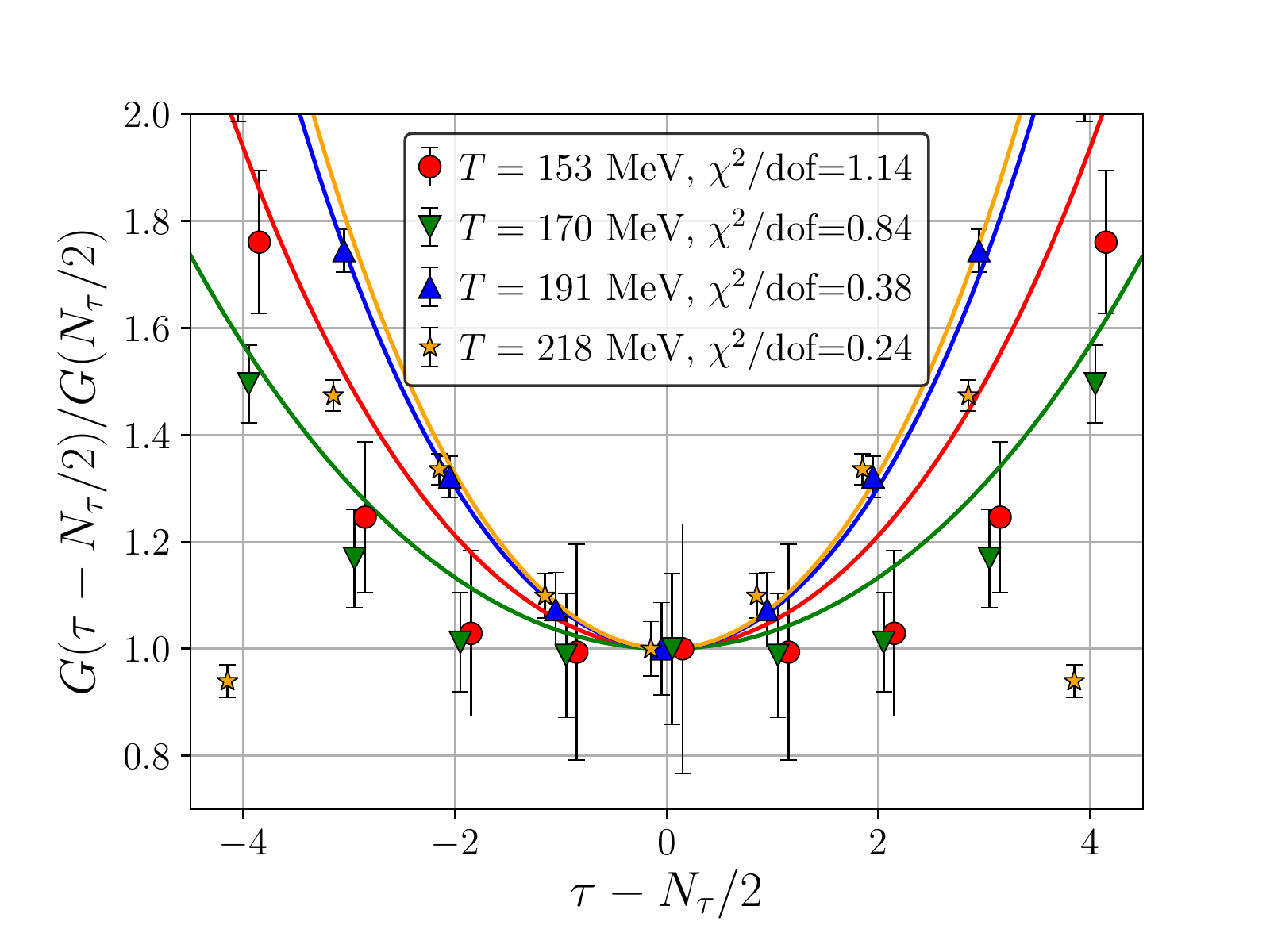}
\end{center}
\caption{Correlators as a function of Euclidean time, shifted and normalized with hyperbolic fits superimposed for pion mass 210 (370)~MeV, left (right). 
The fit extrema are $(0:4]$, with the exception of the two largest
temperatures for pion mass $m_\pi=370$~MeV, which have been fitted in the $(0:3]$ and $(0:2]$ interval, respectively. In the legend the $\chi^2/\text{dof}$ of the fit is also presented.
GF time is $t/a^2=1.5$.
The full details on the results can be seen in Figs.~\ref{fig:masstdep370}--\ref{fig:masstdep210}. 
In order to avoid a complete superposition of data points belonging to different temperatures we applied a tiny shift along the $\tau$ axis.}
\label{fig:variousfits}
\end{figure*}

In short, the behaviour of the $\eta'$ at finite temperature has both
theoretical and experimental significance. However,  
the present understanding of
the behaviour of $\eta{'}$ at high temperatures is incomplete. 
Studies have been carried out~\cite{Torrieri:2018rnx,Horvatic:2018ztu,Gu:2018swy,Nicola:2018vug,Ishii:2016dln,Mitter:2013fxa,Sakai:2013nba,Costa:2008dp}
within the framework of effective field theory, and we will briefly review them
below. The results are inconclusive and call for a first principle analysis.

In this note we will present the first lattice results on the behaviour
of the $\eta'$ around and above the chiral transition. In the paper the finite temperature $\eta'$ mass is defined as the peak of the finite temperature spectral function of a corresponding correlator. Further study and discussions of the spectral function and its shape are to be found later.

\section{$\eta'$, QCD symmetries and topology} 

At the classical level massless QCD enjoys the symmetry $U(N_f) \times U(N_f)
= SU(N_f)_L \times SU(N_f)_R \times U(1)_A \times U(1)_B$. 
In the limit of $N_f=3$ massless quarks
in the normal phase of QCD $SU(3)_L \times SU(3)_R$ is spontaneously broken  
down to $SU(3)_V$, producing an octet of Goldstone bosons, $\pi's$, $K's$,
and $\eta$. 

The associate quark
condensate is not invariant under $U(1)_A$, which, again, is exact at classical
level. 
More precisely, an anomalous $U(1)_A$ transformation would generate a
mass-independent, parity violating term
proportional to $F \tilde F$: 
$\partial_{\mu} J_5^{\mu}
 = m \bar q \gamma_5 q + \frac{1}{32 \pi^2} F \tilde F$.
However, as this term is irrelevant at all orders
in perturbation
theory, it was initially ignored, 
leading to the conclusion that $U(1)_A$ is 
exact in the chiral limit. Since the  quark
condensate, generated by the breaking of
$SU(3)_L \times SU(3)_R$ is not invariant under $U(1)_A$, 
the axial symmetry would
also be spontaneously broken, producing  a ninth, flavor
singlet Goldstone boson,  the $\eta'$.

Physical quark masses
explicitly break chiral symmetry, and the masses of the (now) 
pseudo-Goldstone bosons
may be computed in chiral perturbation theory,
using $\Lambda_{QCD}$ as a sole input,
leading to the well known results: $m_\pi^2 \propto (m_u + m_d) \Lambda_{QCD}$,
$m_K^2 \propto (m_s + m_{u,d}) \Lambda_{QCD}$, 
$m_\eta^2 \propto (m_u + m_d + 4 m_s) \Lambda_{QCD}$. If $U(1)_A$ were
spontaneously broken, the $\eta'$ mass would be in the same mass range.
Its experimental value, $m_\eta'\sim O(1\text{ GeV})$ 
is then incompatible with the chiral perturbation theory result.

\begin{figure*}[hb]
\begin{center}
\includegraphics[width=6cm]{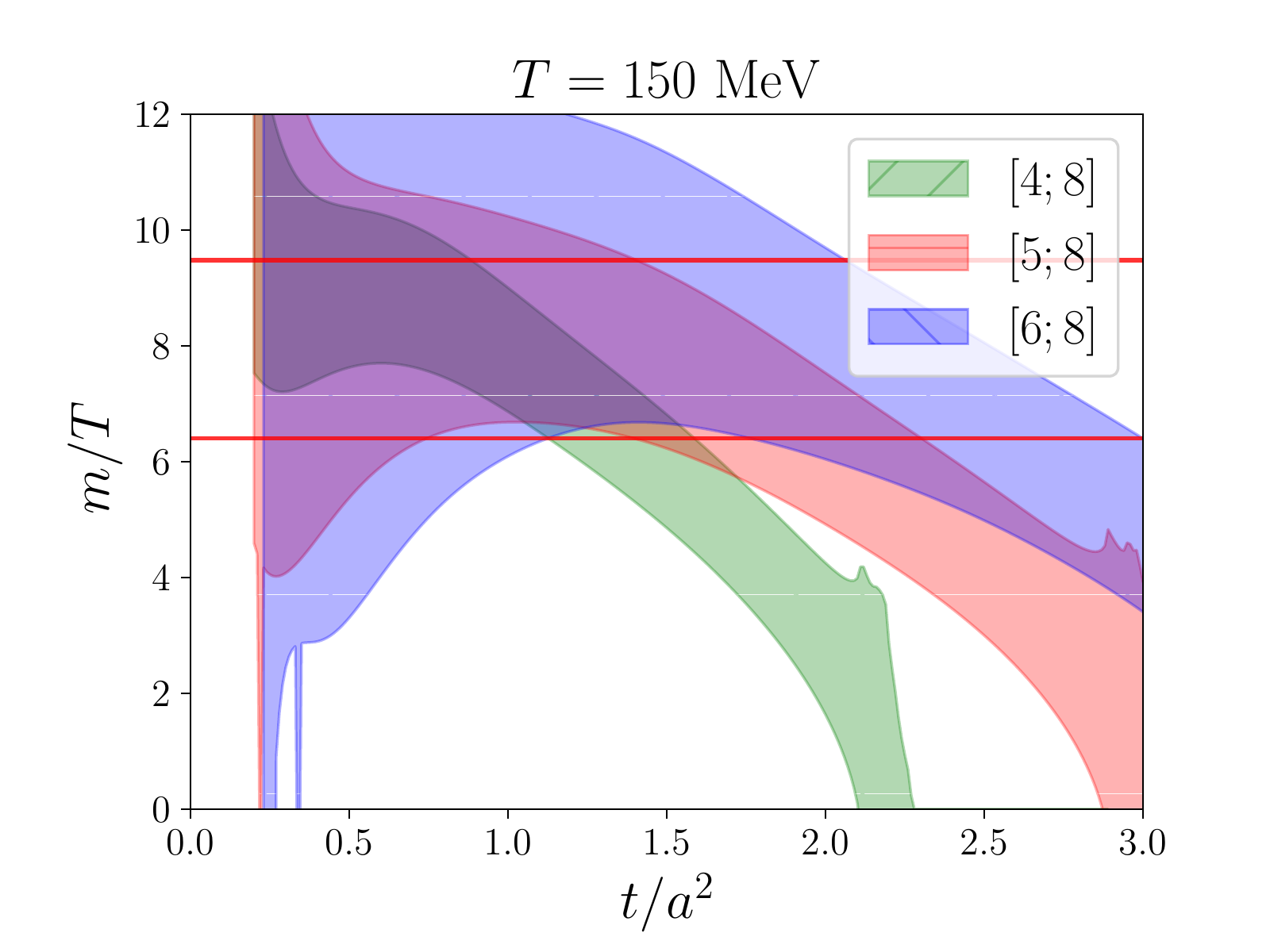}
\includegraphics[width=6cm]{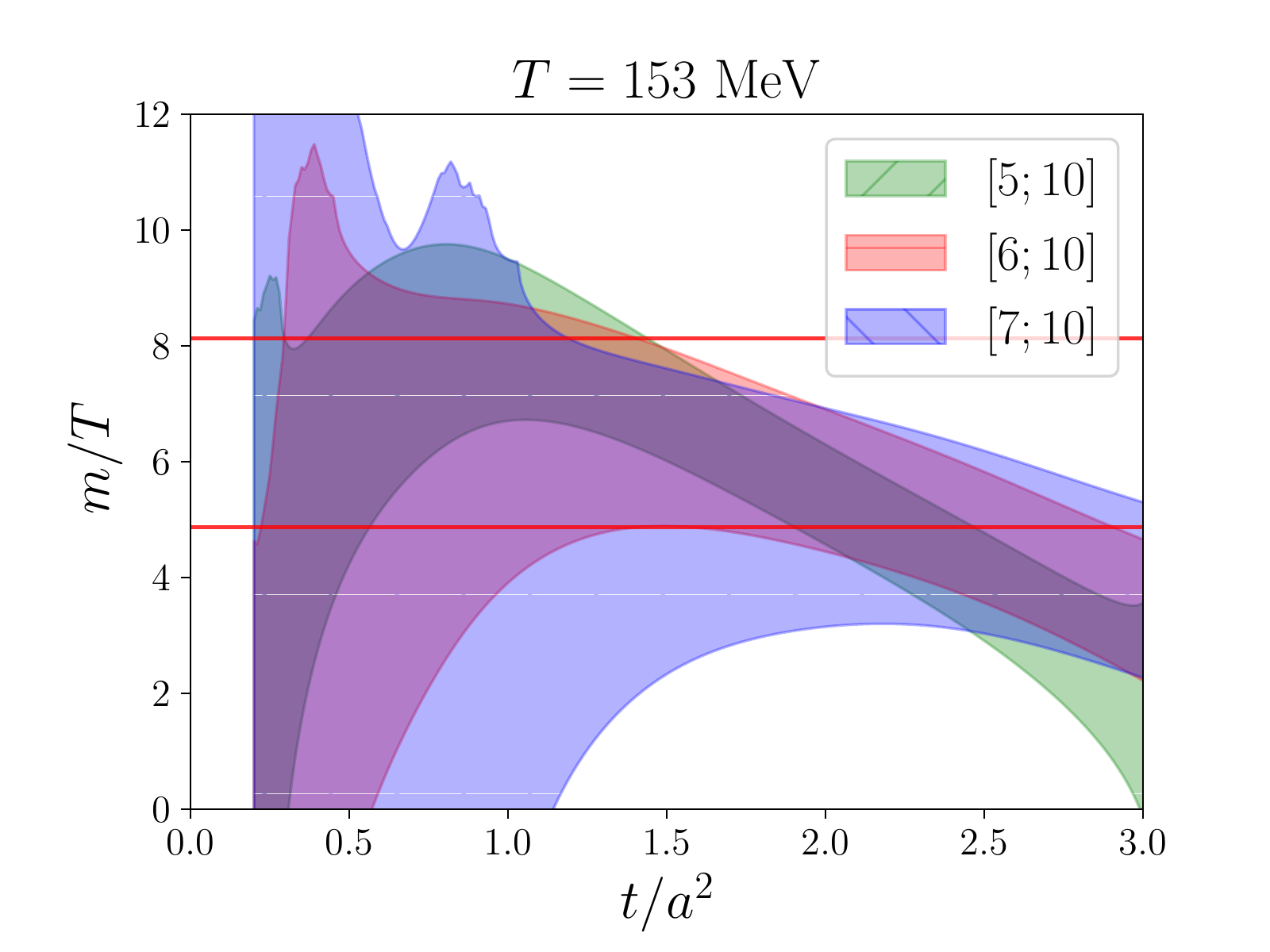}
\includegraphics[width=6cm]{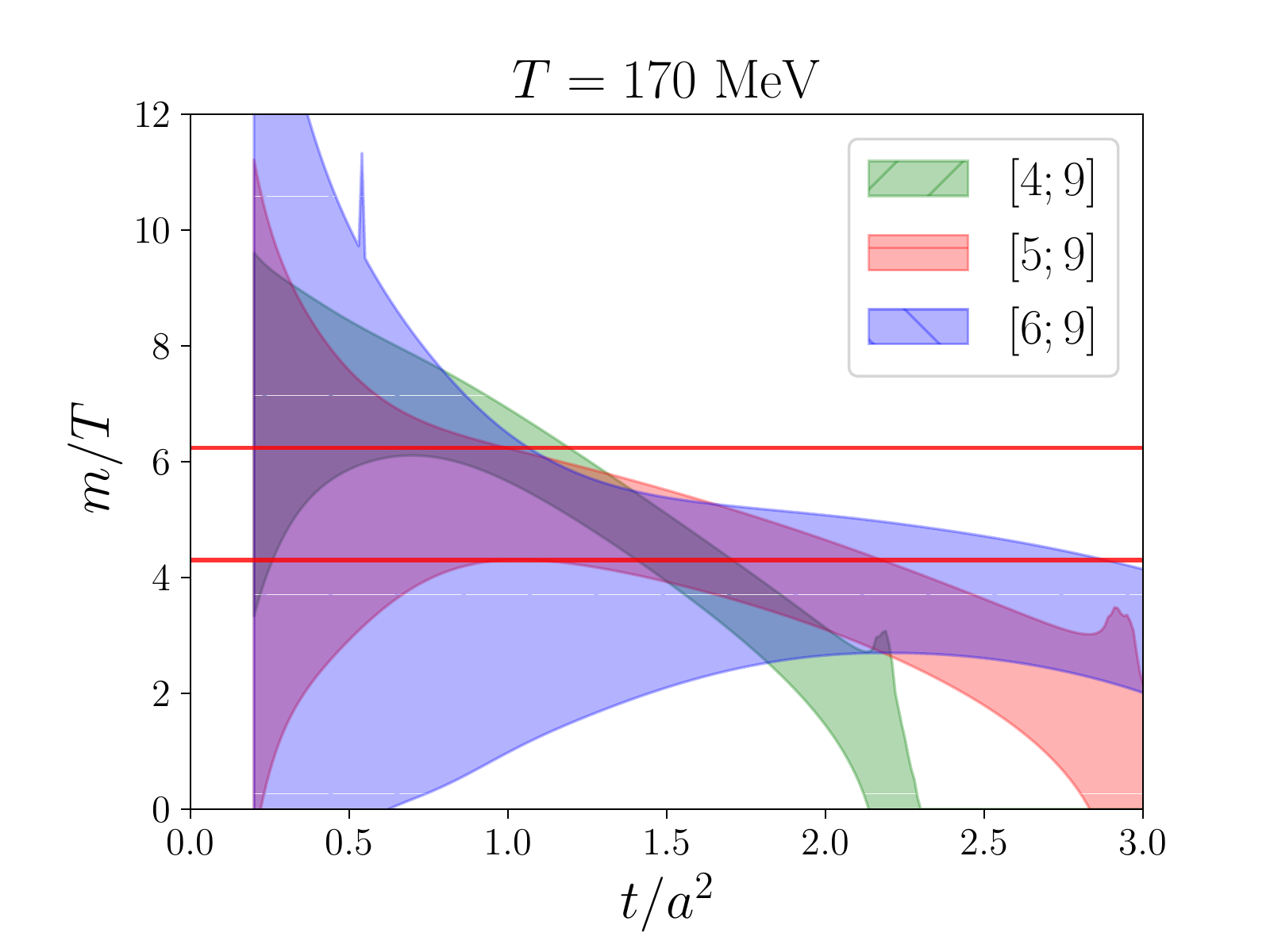}
\includegraphics[width=6cm]{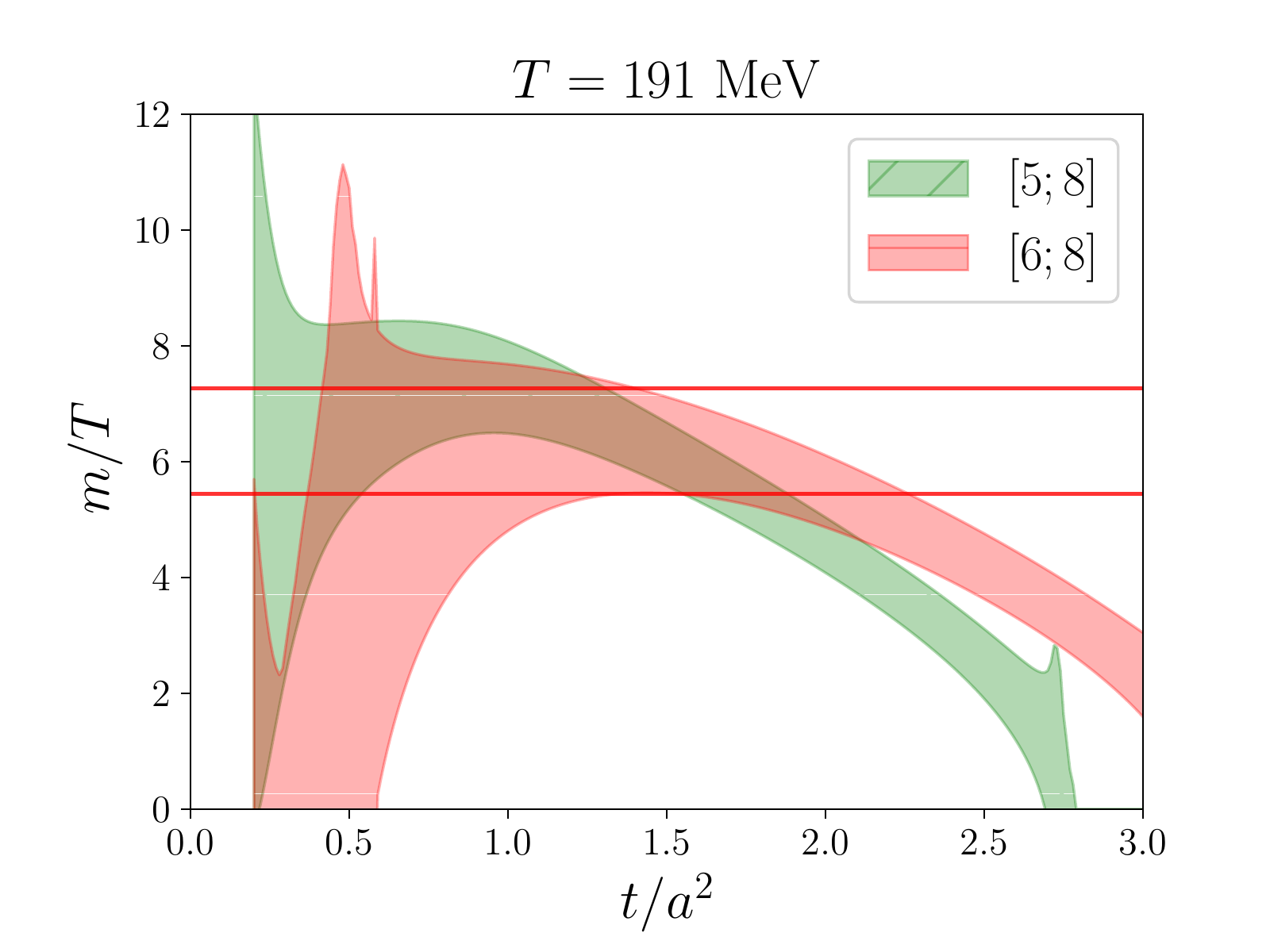}
\includegraphics[width=6cm]{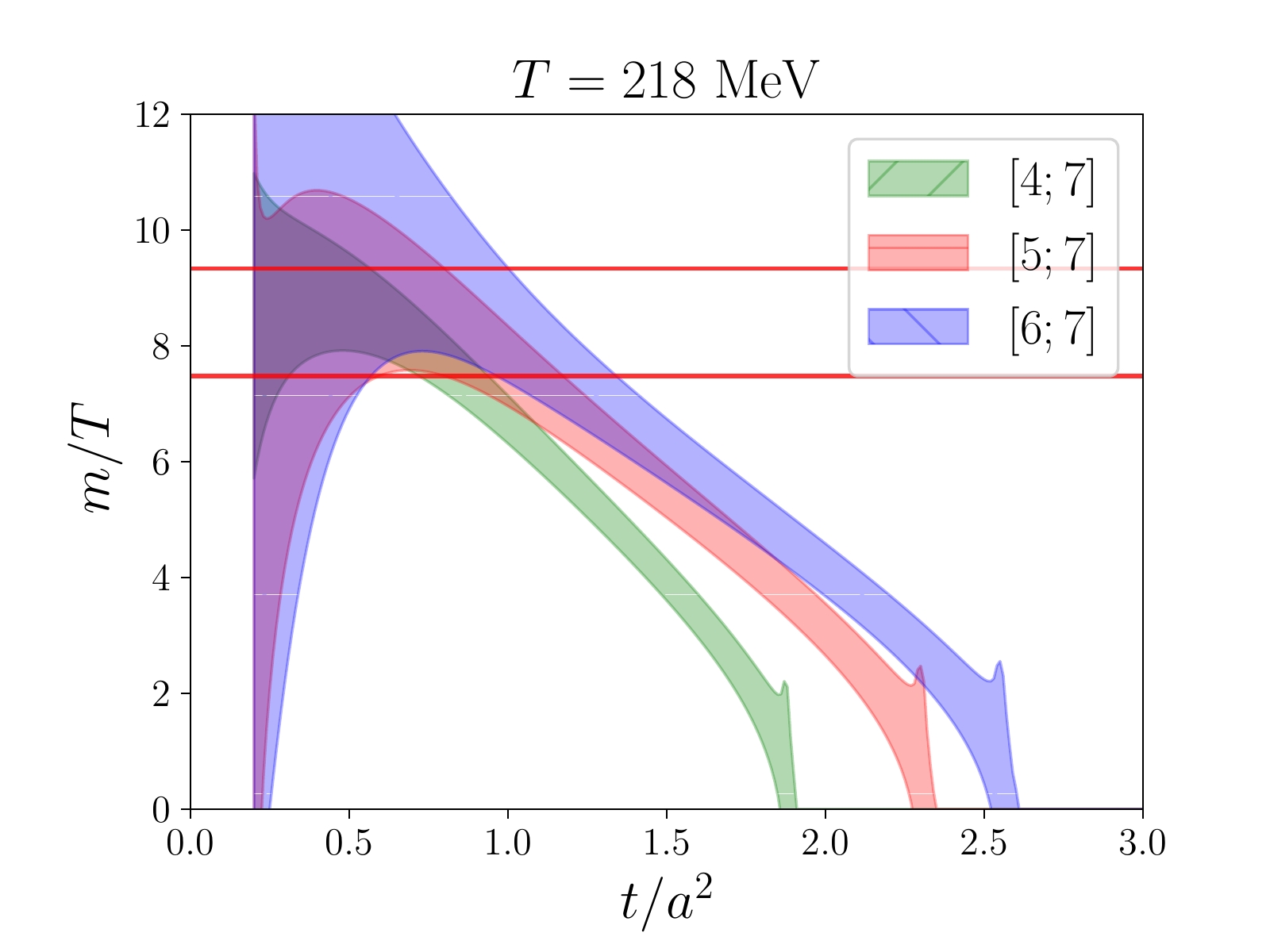}
\caption{Mass extracted from hyperbolic fit within various ranges of Euclidean time $[\tau_{st}, N_\tau/2]$ vs GF time. Lattice step is $a=0.0646(26)$ fm for all plots except for the upper left plot which is for $a=0.0823(37)$ fm. Red horizontal lines indicate the plateau, which is used to determine the mass. Pion mass is $m_{\pi}=370$ MeV.}
\label{fig:masstdep370}
\end{center}
\end{figure*}
\begin{figure*}[b]
\begin{center}
\includegraphics[width=6cm]{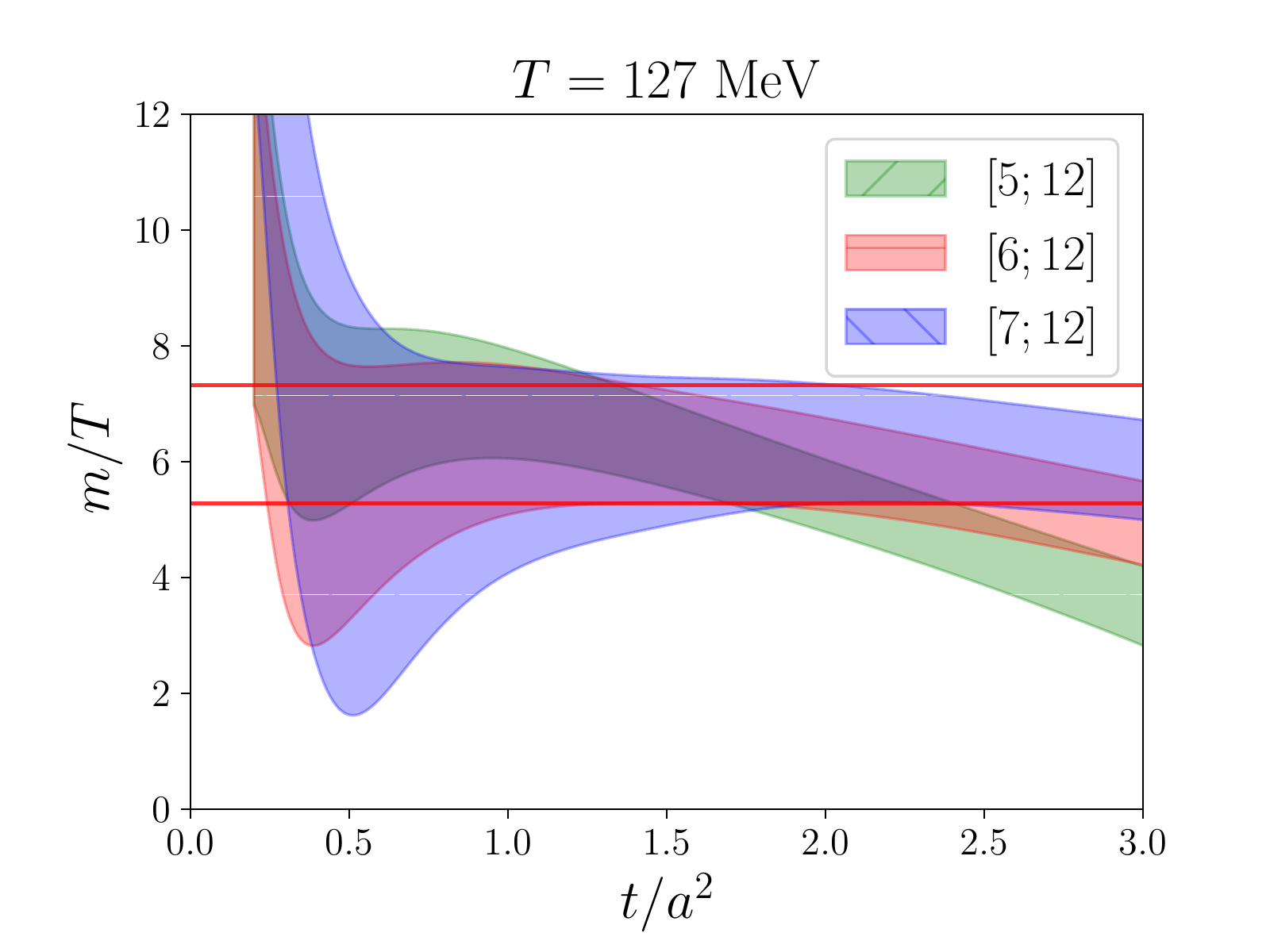}
\includegraphics[width=6cm]{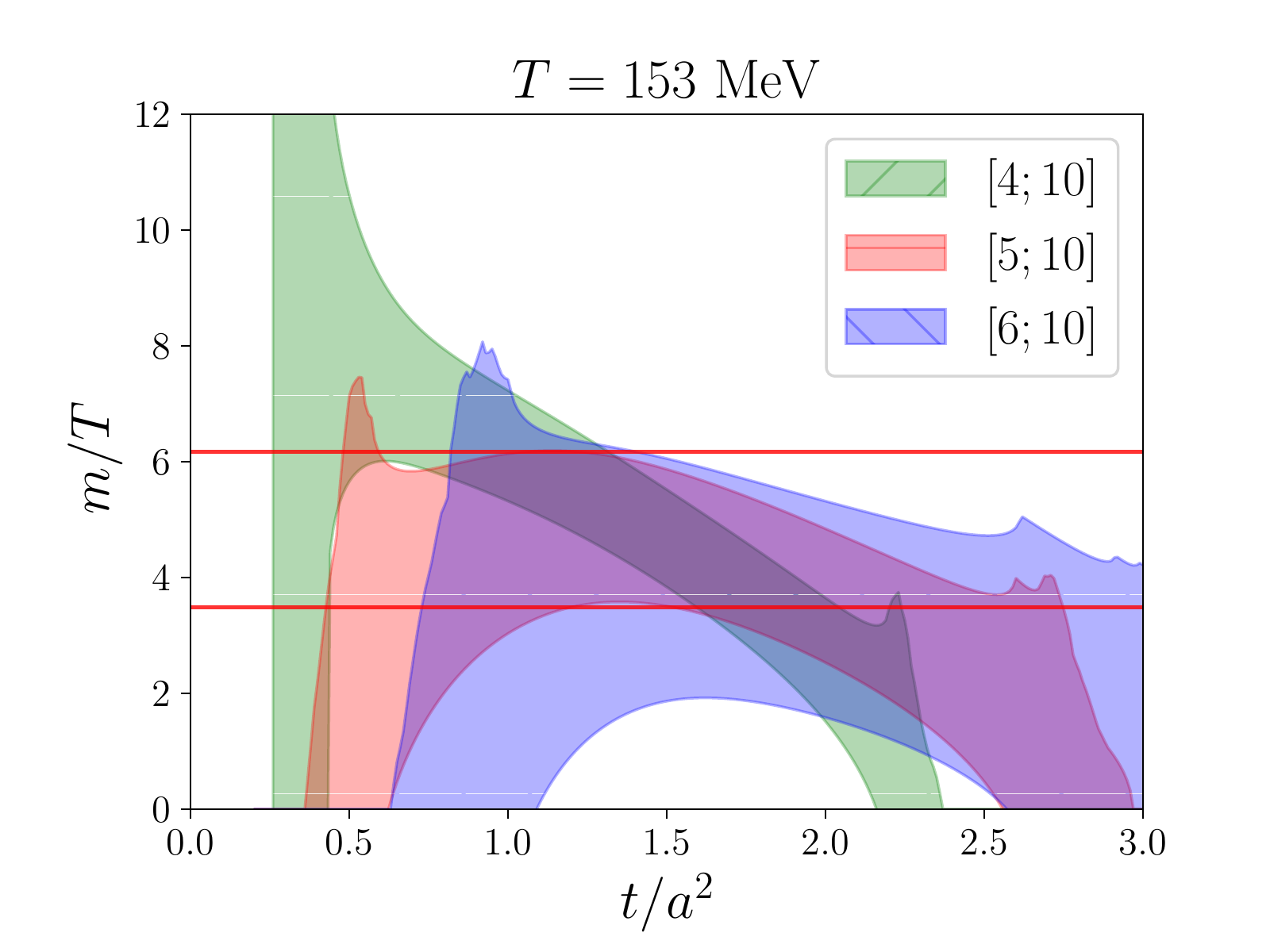}
\includegraphics[width=6cm]{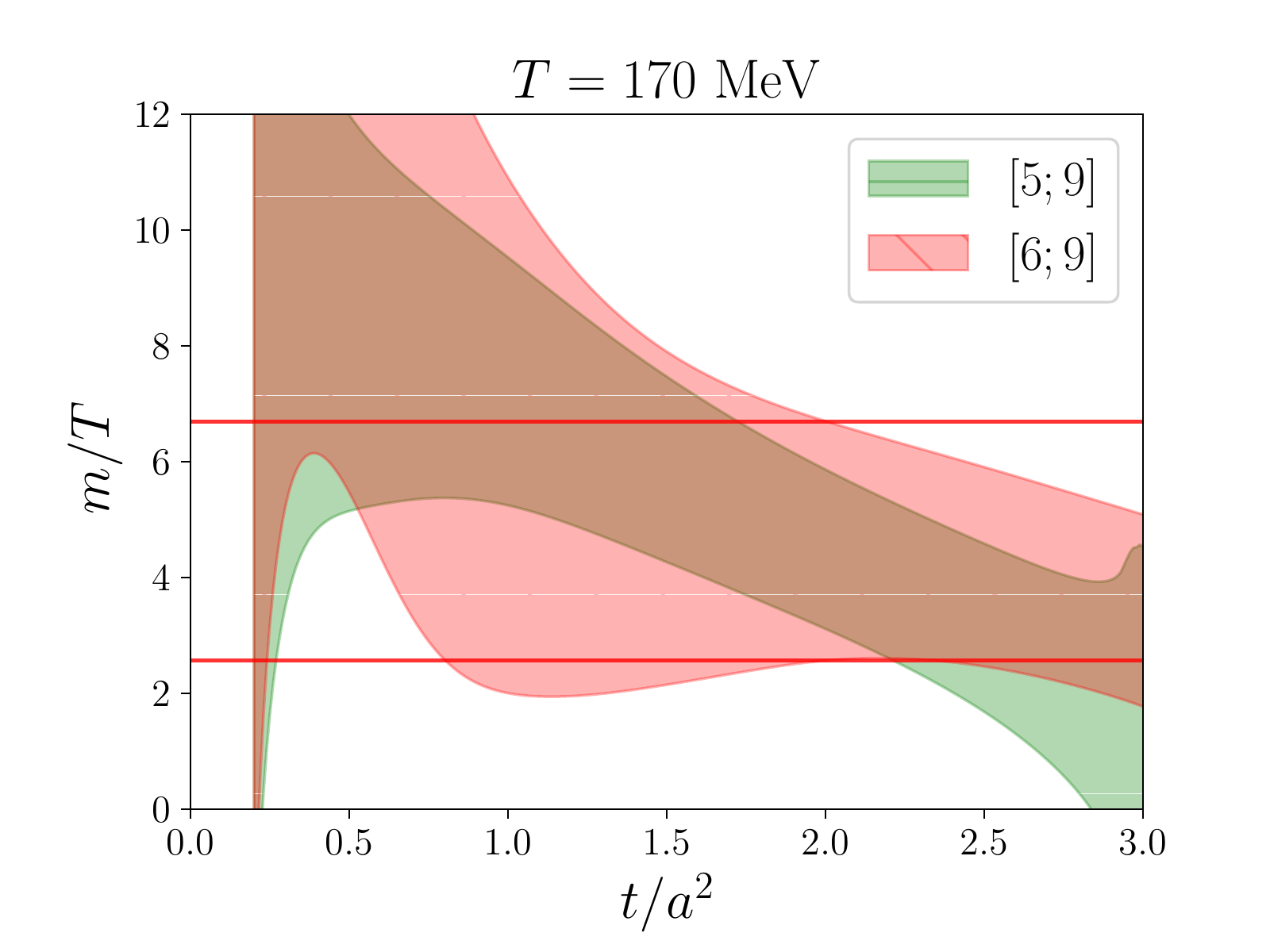}
\caption{Same as Fig. 2, but for $m_{\pi}=210$ MeV. Lattice step is $a=0.0646(26)$ fm.}
\label{fig:masstdep210}
\end{center}
\end{figure*}

As mentioned above, the solution was found
once  realized that $q=\frac{1}{32\pi^2}\tilde FF$ has a topological
nature, hence in principle it could be non-zero:
$\int \tilde FF = 32 \pi^2 Q$, with $Q$ being the topological
charge. Within the framework of 
large-$N$ QCD~\cite{Veneziano:1979ec,DiVecchia:1980yfw,Kawarabayashi:1980dp}
the mass matrix  should then include
an anomalous term $m_A^2=4 N_f/3 f_0^2 \,\chi$, with $\chi = \frac{\langle Q^2\rangle}{V}$ 
being the large-$N$, or, equivalently, the Yang-Mills topological susceptibility. 
In the isospin limit, in which pion and $K$ do not mix with anything else,
adopting the basis $I \equiv \frac{1}{\sqrt 2} (u \bar u + d \bar d); S \equiv s \bar s$, the mass matrix describing the $\eta$ complex reads\cite{Veneziano:1979ec}:
$$
\begin{pmatrix}
m_\pi^2 + m_A^2 & m_A^2/\sqrt2 \\
m_A^2/\sqrt2  & 2 m_K^2 - m_\pi^2 +m_A^2/2
\end{pmatrix}
$$
and we recognize that with $m_A=0$ the $\eta'$ is a pure $s \bar s$ state
with mass  $m_{NA} \simeq 700$ MeV, 
independent on the pion mass. 
Alternatively, $\eta$ and $\eta'$ may be expressed in terms of 
octet-singlet states $\ket{1}=\frac{1}{\sqrt{3}}(u\bar u + d\bar d + s \bar s), \ket{8}=\frac{1}{\sqrt{6}}(u\bar u + d\bar d - 2 s \bar s)$ and a mixing angle $\theta$:
$\ket{\eta} = \cos\theta~ \ket{8} - \sin\theta~ \ket{1}$,
$\ket{\eta'} = \sin\theta~ \ket{8} + \cos\theta~ \ket{1}$.
$\theta$ physical value is about $-20^o$ and would reduce to about
$-50^o$ for $m_A=0$. The mixing is then a useful diagnostic for the
anomalous contribution.  

Beyond leading order in $1/N$, 
there are studies which extend the DGMOR relations to the $U(1)_A$ sector,
however a precise quantitative analysis requires the knowledge of still poorly
known decay constants~\cite{Shore:2007yn}. 

A completely ab initio calculation requires a lattice study.
The standard way to investigate the meson masses in lattice simulations is based on the measurement
of the correlators of corresponding bilinear quark operators $A(x)=\bar{\psi}\Gamma\psi(x)$, where $\Gamma$ acts on flavour and Dirac indices and depends on the meson channel under study.
The $\eta'$ mass may be computed from the decay of
the correlator 
$G^f_{xy}=\langle\bar{\psi}^f\gamma_5\psi^f(x)~\bar{\psi}^f\gamma_5 \psi^f(y)
\rangle$,
however, applying such method for extraction of the $\eta'$ mass poses some complications~\cite{Michael:2013gka,Kaneko:2009za,Christ:2010dd,Gregory:2011sg,Ottnad:2015hva}.
Since $\eta$ and $\eta'$ are not exactly flavour eigenstates, but a mixing of flavour octet and singlet states, one has to analyze the $2 \times 2$ matrix of fermionic current correlators, leading to the masses as well as to the $\eta, \eta'$ mixing angle. Another complication comes from the fact, that the fermionic correlator corresponding to $\eta/\eta'$ channel has also a large so-called disconnected contribution, which determination requires considerable numerical effort.
Alternatively, one may consider  
$G^q_{xy}=1/N^2 \langle q(x) q(y) \rangle $, 
where $q(x)$ is the topological charge density. The correlator
of the topological charge density only contains the $\eta'$
and simply reduces to the Witten-Veneziano formula at leading order. 
At zero temperature with this method the $\eta'$ 
mass was determined in~\cite{Fukaya:2015ara}. 
A recent paper has successfully cross checked the two methods at zero temperature for $N_f=2$ 
twisted mass Wilson fermions~\cite{Dimopoulos:2018xkm}.
At finite temperature this approach was used for extraction of the sphaleron transition rate in gluodynamics~\cite{Kotov:2018vyl}.

\begin{figure*}[t]
\begin{center}
\includegraphics[width=9cm]{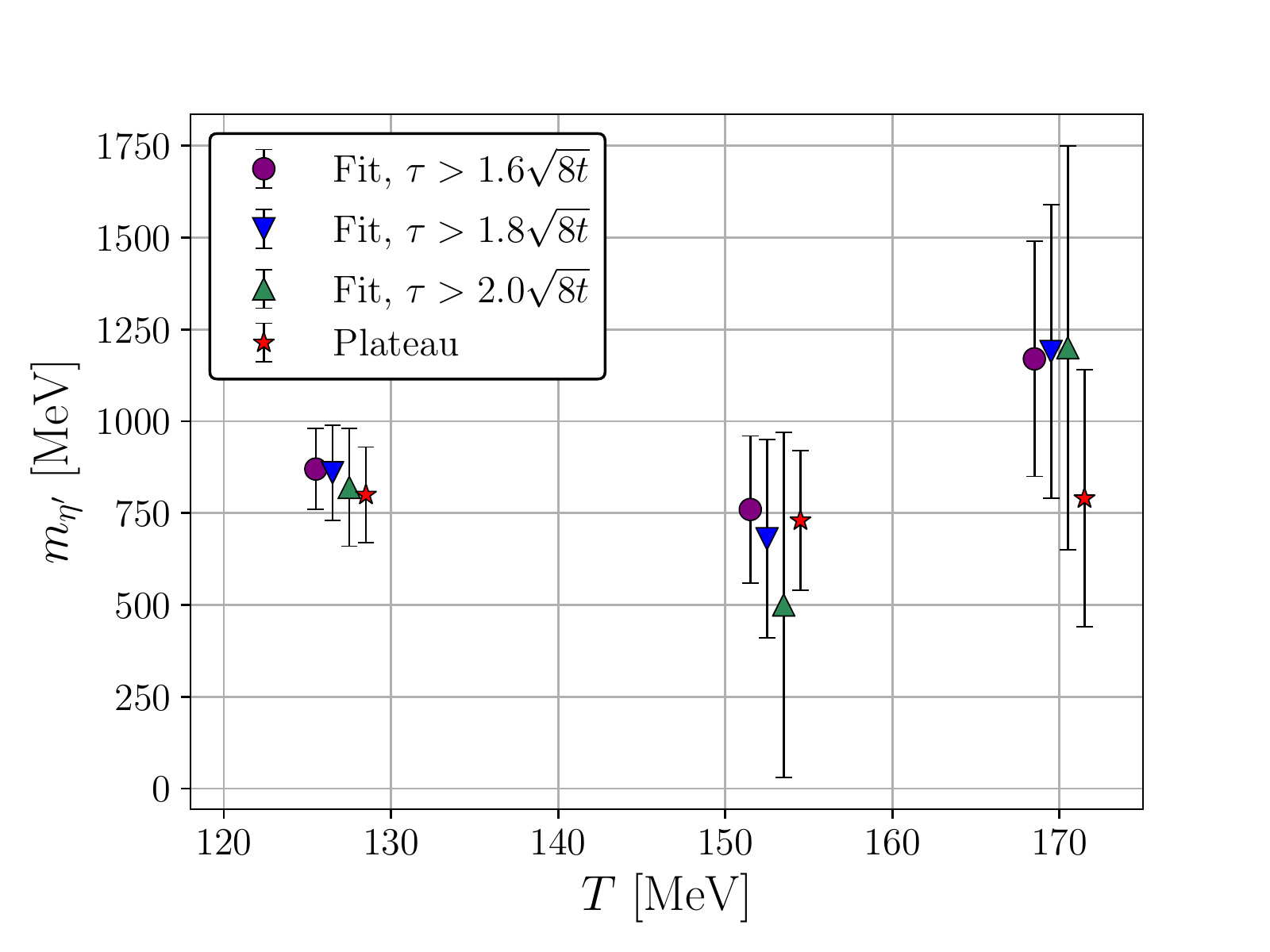}
\includegraphics[width=9cm]{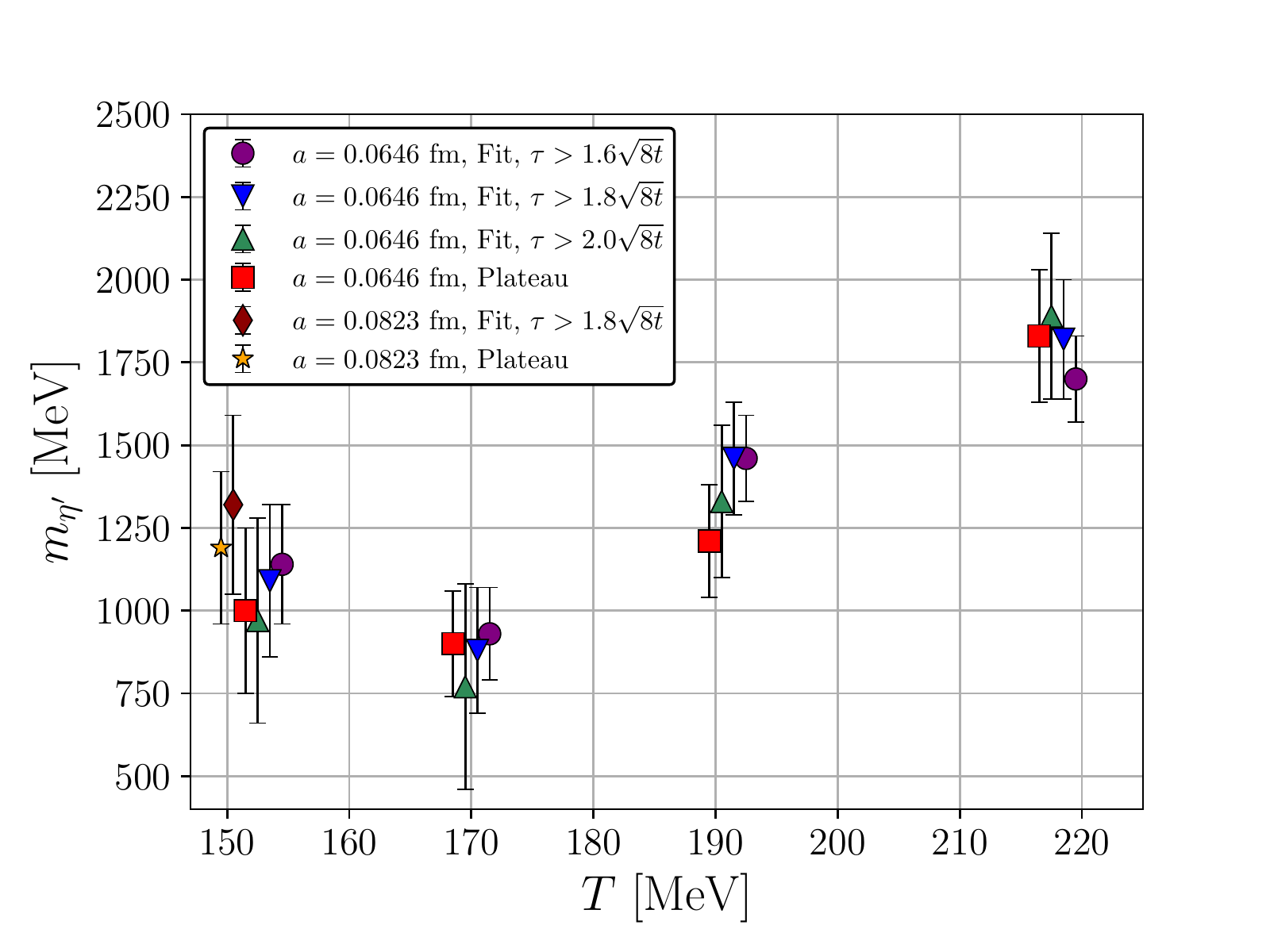}
\end{center}
\caption{Dependence of the $\eta'$ mass extracted by various methods on temperature $T$ for  pion mass 210 (370)~MeV, left (right).
In order to avoid a complete superposition of data points corresponding to different methods we applied a tiny shift along the $T$ axis.}
\label{fig:massartifacts}
\end{figure*}

All in all, at zero temperature, in the broken phase of QCD, we have
a quantitative understanding of the $\eta$ mass complex. 
At finite temperature the situation is far less clear. 
Already by looking at the basic relations, it is clear that the fate
of the $\eta'$ at high temperature is entangled with the behaviour
of the light mesons as well as with that of the topological susceptibility.
In symmetry languages, with chiral
and axial symmetries restoration.  

Chiral and $U(1)_A$ restoration at high temperatures and
their implications on the $\eta$ mesons  have been 
considered by several authors~\cite{Torrieri:2018rnx,Horvatic:2018ztu,Gu:2018swy,Nicola:2018vug,Ishii:2016dln,Mitter:2013fxa,Sakai:2013nba,Costa:2008dp}.
A generalization of the Witten-Veneziano
approach at finite temperature~\cite{Horvatic:2018ztu} 
indicates a dip in the $\eta'$ mass
followed by an increase, leading to a near-degeneracy with the
pion mass. This approach requires a parameterization of the temperature
dependence of the anomalous contribution to the mass, which has a strong
quantitative influence on the results, even if some general features~--
the dip, and the increase -- are robust. Ref.~\cite{Nicola:2018vug} 
used Ward identities and $U(3)$ ChPT to study the pattern of degeneration
of chiral partners and mixing angles. Also in this study the $\eta'$ mass
has a drop, and the mixing angle approaches the ideal one, consistent
with the disappearance of the anomaly and the restoration of the 
$SU(2) \times SU(2) \times U(1)_A$ symmetry. Analogous results were found
within a Polyakov Loop extended NJL model~\cite{Ishii:2016dln}. An earlier
interesting work~\cite{Mitter:2013fxa}  finds again a reduction
of the $\eta'$ consistent with experiments, within an exact functional 
renormalization group approach, discussing in detail the role of the anomaly,
but only within the $SU(3) \times SU(3)$ symmetry. 

These studies show some qualitative common features, but differ in quantitative
details. 
In the following we are going to present our lattice results for the temperature
dependence of the $\eta'$ mass. 

\begin{table}
\begin{center}
\begin{tabular}{|c|c|c|c|c|c|}
\hline
 $m_{\pi}$ [MeV] & $N_{\tau}\times N_{\sigma}^3$ & $a$ [fm] & $T$ [MeV] & \
\
\# conf\\
\hline
369 & $20\times 48^3$ & 0.0646(26) & 153(6) & 1173\\
369 & $18\times 40^3$ & 0.0646(26) & 170(7) & 1198\\
369 & $16\times 32^3$ & 0.0646(26) & 191(8) & 3879\\
369 & $14\times 32^3$ & 0.0646(26) & 218(9) & 3965\\
\hline
372 & $16\times 32^3$ & 0.0823(37) & 150(7) & 2679\\
\hline
213 & $24\times 48^3$ & 0.0646(26) & 127(5) & 1120\\
213 & $20\times 48^3$ & 0.0646(26) & 153(6) & 552\\
213 & $18\times 48^3$ & 0.0646(26) & 170(7) & 470\\
\hline
\end{tabular}
\end{center}
\caption{Parameters and statistics used in the simulations.}
\label{tab:parameters}
\end{table}

\section{Lattice details and methodology}
\label{sec:lattice}
The results have been obtained on lattice configurations with $N_f=2+1+1$
flavors of twisted mass Wilson fermions, on a subset of parameters 
used in~\cite{Burger:2018fvb}. In some cases we have enlarged the statistics
by performing new simulations. The mass 
parameters of $s$ and $c$ quarks are fixed at their physical values and
several sets of light doublet masses are available, all corresponding to 
larger  than physical values of pion mass~$m_\pi$.
In the following we use two values of pion mass, $m_{\pi}\simeq210$~MeV and $m_{\pi}\simeq370$~MeV.
In Tab.~\ref{tab:parameters} the parameters and statistics used in our 
simulations are summarized, while in Tab.~\ref{tab:Tc} we note,
for future reference, the pseudocritical temperatures for the two pion 
masses considered in this work. 
Let us remind the reader that for a non-zero quark mass the transition
turns into a crossover, whose position is prescription dependent: 
in Tab.~\ref{tab:Tc} we 
show the crossover temperatures associated with the maximum
of the derivative with respect to the mass and 
with the maximum of the derivative w.r.t. the temperature  
of the chiral order parameter, denoted as $T_\chi$ and $T_\Delta$. 

\begin{table}
\begin{center}
  \begin{tabular}{|c|c|c|c|}
\hline
$m_\pi$ [MeV]& $a$ [fm] & $T_\chi$ [MeV] & $T_\Delta$ [MeV] \\
 \hline
369 & 0.0646(26) & 185(1)(3)  & 180(5)(1)  \\
372 & 0.0823(37) & 189(2)(1)  & 194(2)(0)  \\
213 & 0.0646(26) & 158(1)(4)  & 165(3)(1)  \\
\hline
\end{tabular}
\end{center}
\caption{The pseudo-critical temperatures associated
with the mass (temperature) derivatives of the chiral order parameters.
The first error is statistical, the second systematic, see text for details.
Adapted from Ref.~\cite{Burger:2018fvb}.}
\label{tab:Tc}
\end{table}

\begin{figure*}[b]
\begin{center}
\includegraphics[width=12cm]{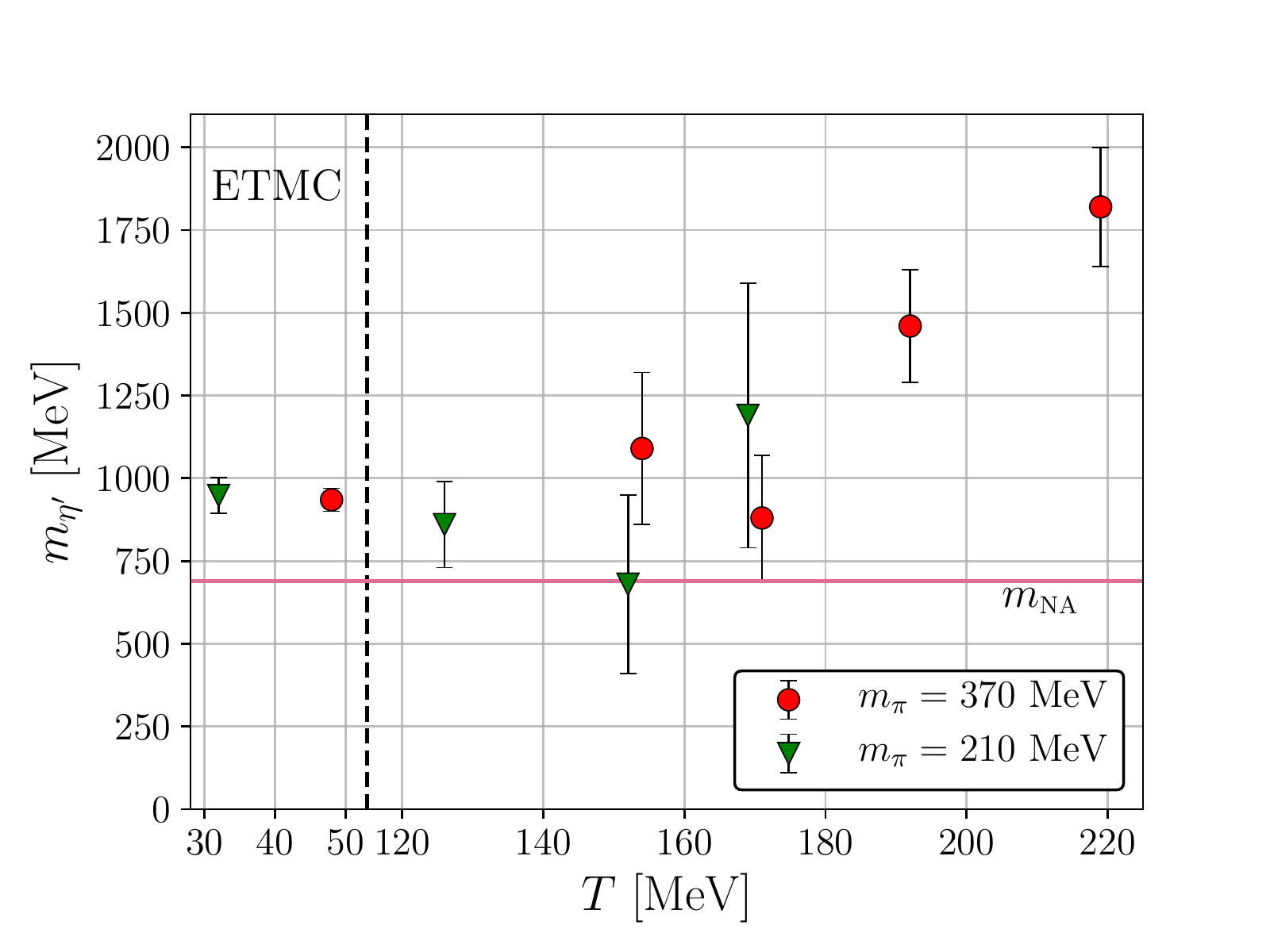}
\end{center}
\caption{Dependence of the extracted value of the $\eta'$ mass on the temperature for both pion masses \
$m_{\pi}=210$ MeV and $370$ MeV under study. Also, the low-temperature results of~\cite{Ottnad:2017bjt} are shown for comparison. Horizontal line indicates the zero temperature non-anomalous 
contribution $m_{\mbox\small NA}\approx 700$ MeV to the $\eta'$ mass.
In order to avoid a complete superposition of data points belonging to different pion masses we applied\
 a tiny shift along the $T$ axis.}
\label{fig:finalres}
\end{figure*}
\begin{figure*}[t]
\begin{center}
\includegraphics[width=12cm]{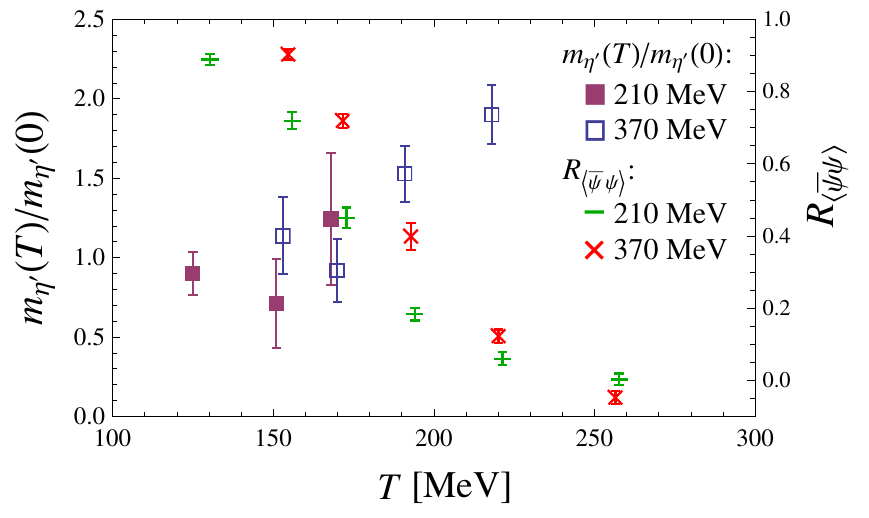}
\end{center}
\caption{The ratio of the temperature-dependent mass $m_{\eta'}(T)$ to its zero-temperature
value $m_{\eta'}(0)$ superimposed with the results for the renormalized chiral condensate. $a=0.0646(26)$~fm. In order to avoid a complete superposition of data points belonging to different pion masses we applied\
 a tiny shift along the $T$ axis.}
\label{fig:mass_top_1}
\end{figure*}
The mass of the $\eta'$ is extracted from the correlator of the topological charge density:
\begin{equation}\begin{split}
	G(\tau)=\int d^3\bar{x} q(0)q(\tau,\bar{x}) = \int d^3\bar{x}\frac{1}{32\pi^2} & F_{\mu\nu}\tilde{F}_{\mu\nu}(0)\times\\
	\times&\frac{1}{32\pi^2}F_{\mu\nu}\tilde{F}_{\mu\nu}(\tau,\bar{x})
\label{eq:correlator}
\end{split}\end{equation}
To measure the topological charge density correlator we used (Wilson) Gradient Flow (GF)~\cite{Luscher:2010iy}. Starting from lattice gauge field $U(x,\mu)$, the evolution of the gauge variables in auxiliary GF time $t$ is performed:
\begin{equation}
\partial_t V_t(x,\mu) = - \partial_{x,\mu} \Bigl\{ S_W[V_t] \Bigr\} V_{t}(x,\mu), \quad V_0(x,\mu)= U(x,\mu).
\end{equation}
This effectively smears the gauge fields over the range $\sim\sqrt{8t}$. After performing GF, we measured the correlator of the topological charge density $q(\tau,\bar{x})=\frac{1}{32\pi^2}\epsilon_{\mu\nu\rho\sigma}F_{\mu\nu}F_{\rho\sigma}(\tau,\bar{x})$ as a function of GF time based on field variables $V_t$. For field strength tensor $F_{\mu\nu}$ we used a clover discretization. 

	In \cite{Trunin:2015yda} such lattice definition of topological charge density was used for studying the topological susceptibility on the same ensembles. The results of \cite{Trunin:2015yda} indicate that all relevant topological sectors are correctly sampled without topological freezing.

Since Gradient Flow smears the gauge fields, at large GF times $t$ the contribution of higher excited states to the correlator $G(t,\tau)$ become suppressed and the statistical noise in $G(t,\tau)$ is reduced.

However, the correlator~\eqref{eq:correlator} has a large positive contact term at $C(\tau=0)$\footnote{Note that the correlator~\eqref{eq:correlator} is negative at $\tau\ne0$ due to pseudoscalar nature of the topological charge density $q$.}, which is not related to any propagating physical degree of freedom. At GF times $t\gtrsim\tau^2/8$ the smeared contact term significantly changes the correlator $G(t,\tau)$ and such points are omitted in our analysis.

Within the discussed above restrictions the correlator $G(t,\tau)$ should have a simple behaviour $G(t,\tau)\sim \cosh\left\{m(\tau-N_{\tau}/2)\right\}$, with the parameter $m$ giving the mass of the $\eta'$.
Our discussion ignores the width of the spectral function: we will
see that the data are compatible with the simple spectral decomposition
we have assumed. This indicates that, within our errors, the spectral function
remains narrow enough to allow the identification of the mass of the
ground state (or, more precisely, with an average value of the mass 
range spanned by the peak).

\section{Results}

To have a first feeling of the results, we select a fixed value of the GF time $t/a^2=1.5$ 
and in Fig.~\ref{fig:variousfits} we plot the normalized correlator $G(\tau-N_{\tau}/2)/G(N_{\tau}/2)$ for various temperatures, at 
$a=0.0646(26)$ fm. We superimpose hyperbolic fits $\cosh\left\{m(\tau-N_{\tau}/2)\right\}$ (the systematic thereof will
be discussed momentarily): for the lighter pion mass the data suggest a small drop
around $T=150$ MeV, followed by an increase of the mass, for a pion mass of about
370 MeV the qualitative trend is similar, but the drop happens around 
$T = 170$ MeV. We also present the $\chi^2/\text{dof}$ of the fit in the Fig.~\ref{fig:variousfits}. Note that here we use the whole covariance matrix. Since $\chi^2/\text{dof}\sim 1$, we can state, that the correlator within the statistical uncertainty
can be described by simple $\cosh$-like form.

Let us go further and analyze the GF time dependence of the correlator. 
For each GF time $t$ we fit the correlator $G(\tau)$ with hyperbolic fit:
$G(\tau)\sim A\cosh\left\{m(\tau-N_{\tau}/2)\right\}$ within the range $\tau\in[\tau_{st}, N_\tau/2]$, where we changed the left point $\tau_{st}$ of the fit range. We plot the result of the fit for $m(t)$ as a function of the GF time $t$ for several fitting ranges $\tau\in[\tau_{st}, N_\tau/2]$ in Figs.~\ref{fig:masstdep370}--\ref{fig:masstdep210}. The plateau in the $t$-dependence corresponds to the $\eta'$ mass. We find the smallest value of the left end $\tau_{st}$ of the interval $[\tau_{st}, N_\tau/2]$, which still has a plateau and by the height of this plateau extracted the mass. The results are shown by horizontal lines in Figs.~\ref{fig:masstdep370}--\ref{fig:masstdep210} and summarized in the fourth column of the Tab.~\ref{tab:results}.

As an alternative method, we fit the correlator $G(t,\tau)$ simultaneously for all points  with the restriction: $\tau/a\ge4$, $\tau>b\sqrt{8t}$ by the function:
\begin{equation}
G(t,\tau) = a(t)\cosh m(\tau-N_\tau/2),
\end{equation}
Note that $m$ is the same for all values of GF time $t$, while the coefficient $a(t)$, in principle, depends on $t$. By varying the number $b$ we change the region of the fitting, thus we can balance between statistical and systematic error (see the discussion in Sec.~\ref{sec:lattice}). Since the characteristic radius of smearing is $\sim\sqrt{8t}$, one should take $b\sim 2$. In the following we used $b=2$ as well as smaller numbers $b=1.6$ and $b=1.8$.
We present the value of the $\eta'$ mass extracted by this method, along with the mass, extracted by previous method, in Fig.~\ref{fig:massartifacts}. From these figures one sees that the results extracted by various methods, as well as the results for different lattice steps are in agreement with each other. If one increases $b$, that is, decreases the fitting range, the errorbars grow larger, however the results for various ranges  $b$ are in agreement with each other. As the final estimation we take the results of the fit for  the smaller lattice step $a=0.0646(26)$~fm and $b=1.8$, which are also presented in the Tab.~\ref{tab:results}. For comparison, we quote in the Tab.~\ref{tab:results} the results reported in~\cite{Ottnad:2017bjt} for the $\eta'$ mass at low temperatures
for $N_f=2+1+1$ flavors of twisted mass Wilson fermions and the same setup which we have used here.

\begin{table}
\begin{center}
\begin{tabular}{|c|c|c|c|c|}
\hline
$T$, MeV & $\beta$ & $m_\pi$, MeV & $m_{\eta'}$, MeV, & $m_{\eta'}$, MeV,\\
&&&plateau & fit \\
\hline
153(6) & 2.10 & 370 & $1000\pm250$ & $1090\pm230$ \\
170(7) & 2.10 & 370 & $900\pm160$ & $880\pm190$\\
191(8) & 2.10 & 370 & $1210\pm170$ & $1460\pm170$\\
218(9) & 2.10 & 370 & $1830\pm200$ & $1820\pm180$\\
\hline
150(7) & 1.95 & 370 & $1190\pm230$ & $1320\pm270$ \\
\hline 
127(5) & 2.10 & 210 & $800\pm130$ & $860\pm130$ \\
153(6) & 2.10 & 210 & $730\pm190$ & $680\pm270$ \\
170(7) & 2.10 & 210 & $790\pm350$ & $1190\pm400$ \\
\hline
\multicolumn{5}{|c|}{ETMC} \\
\hline
48 & 2.10 & 370 & \multicolumn{2}{|c|}{$935\pm35$}\\
32 & 2.10 & 210 & \multicolumn{2}{|c|}{$948\pm54$}\\
\hline
\end{tabular}
\end{center}
\caption{Values of $\eta'$ mass extracted by plateau analysis (see Fig.~2-3), as well as by single simultaneous fit for all points with $\tau/a\ge4$, $\tau>1.8\sqrt{8t}$. In the last two lines we quote the results of~\cite{Ottnad:2017bjt} at small temperatures.}
\label{tab:results}
\end{table}

\section {Discussion}

The final plot for the dependence of $m_{\eta'}$ as a function of temperature is shown in Fig.~\ref{fig:finalres}. 
In the same plot we show the low temperature result of Ref.~\cite{Ottnad:2017bjt}.
We note that the $\eta'$ mass is rather robust against temperature, with a suggested small dip around the pseudocritical
temperature, followed by an increase at larger temperature.
In the same plot we also indicate, as a horizontal line, the zero temperature non-anomalous component of the $\eta'$ computed at leading order in $1/N$,
$m_{s \bar{s}} \simeq 700$ MeV. Clearly more statistics will be needed to assess  the quantitative
reduction in mass. At the same time, the trend of the propagators gives some confidence in this observation. Indeed, and
interestingly, the reduction, and subsequent increase,
 seems correlated with a signal and the light flavor sector: 
in Fig.~\ref{fig:mass_top_1} we superimpose our results for the $\eta'$ mass with those for the renormalized chiral condensate 
$R_{\langle \bar\psi\psi\rangle}$ obtained on the same lattices in Ref.~\cite{Burger:2018fvb}. Also, 
the pseudocritical temperature associated with the inflection point 
of the chiral condensate, or with the peak of the disconnected 
susceptibility can be
read off Tab.~\ref{tab:Tc}, again suggesting 
a correlation with the region of the dip of the $\eta'$ mass. 
The reduction of the $\eta'$ mass observed at finite temperature
may be due to
a vanishing anomalous contribution, 
which, if confirmed, could indicate that
the ideal mixing has been reached.

In conclusion, we have studied the $\eta'$ for two different
pion masses, and varying temperatures. We have
 observed a correlation between
the behaviour of the $\eta'$ at high temperature 
and that of the chiral observables. 
We confirm  that the $\eta'$  
increases above the pseudocritical temperature associated with 
the light degrees of freedom. The results
are consistent with a modest dip correlated with the same 
pseudocritical temperature. The magnitude of the dip is 
not incompatible with the 200 MeV reduction 
indicated by experimental analysis, and of confirmed would
bring the $\eta'$ mass close to the zero temperature
non-anomalous result, implying ideal mixing in the $\eta$ sector.

These results  are open to different interpretation:
they may be linked with the $SU(2) \times SU(2) \times U(1)_A$ 
restoration, or they may
be simply a manifestation of the complicated $SU(N_f) \times SU(N_f)$,
$N_f = 2$ or $N_f=3$, dynamics in the light sector. 
A more complete set of observables, including the light meson spectrum,
the $\eta$ mass and the mixing angle, is needed in order to draw
conclusive results on these issues.

\section*{Acknowledgments}
It is a pleasure to thank Dubravko Klabučar, 
Frithjof Karsch, Swagato Mukherjee, Angel G{\'o}mez Nicola, Laura Tolos and Giancarlo Rossi for conversations and correspondence.  
The work of A.Yu.K. was supported by grant from the Russian Science Foundation (project number 18-72-00055). 
M.P.L. acknowledges the 
hospitality of the Galileo Galilei Institute for Theoretical
Physics and the support of the European Cost Action 
CA15213 - THOR, Theory of hot matter and relativistic heavy-ion collisions.
A.M.T. acknowledges support from the BASIS foundation.
This work has been partly carried out using computing resources of CINECA (INFN-CINECA agreement)
and the federal collective usage center Complex for Simulation and Data Processing for Mega-science Facilities at NRC ``Kurchatov Institute'', http://ckp.nrcki.ru/.

\bibliographystyle{unsrt}

\end{document}